# COMPARISON OF DOWNSCALING TECHNIQUES FOR HIGH RESOLUTION SOIL MOISTURE MAPPING

*Sabah Sabaghy[1], Jeffrey Walker[1], Luigi Renzullo[2], Ruzbeh Akbar[10], Steven Chan[3], Julian Chaubell[3], Narendra Das[3], R. Scott Dunbar[3], Dara Entekhabi[5], Anouk Gevaert[4], Thomas Jackson[6], Olivier Merlin[7,8], Mahta Moghaddam[10], Jinzheng Peng[11,12], Jeffrey Piepmeier[11], Maria Piles[9], Gerard Portal[13], Christoph Rüdiger[1], Vivien Stefan[7], Xiaoling Wu[1], Nan Ye[1], and Simon Yueh[3]*

[1] Department of Civil Engineering, Monash University, VIC 3800, Australia.
[2] CSIRO Land and Water, Canberra, ACT.
[3] Jet Propulsion Laboratory, California Institute of Technology, Pasadena, CA 91109 USA.
[4] VU University Amsterdam, Department of Earth Sciences, Earth and Climate Cluster, De Boelelaan 1085, 1081 HV Amsterdam, The Netherlands.
[5] Department of Civil and Environmental Engineering, Massachusetts Institute of Technology, Cambridge, MA 02139 USA.
[6] U.S. Department of Agriculture ARS, Hydrology and Remote Sensing Laboratory, Beltsville, MD 20705 USA.
[7] Centre d'Etudes Spatiales de la Biosphère (CESBIO), 18 Avenue, Edouard Belin, bpi 2801, Toulouse 31401, France.
[8] Faculté des Sciences Semlalia Marrakech (FSSM), Avenue Prince Moulay Abdellah, BP 2390, Marrakech 40000, Morocco.
[9] Image Processing Lab (IPL), Universitat de València, Spain.
[10] Department of Electrical Engineering, University of Southern California, Los Angeles, CA 90089 USA.
[11] Goddard Space Flight Center, 8800 Greenbelt Rd, Greenbelt, MD 20771, USA.
[12] Universities Space Research Association, Columbia, MD, USA.
[13] Universitat Politècnica de Catalunya, IEEC/UPC and Barcelona Expert Centre, Spain.

## ABSTRACT

Soil moisture impacts exchanges of water, energy and carbon fluxes between the land surface and the atmosphere. Passive microwave remote sensing at L-band can capture spatial and temporal patterns of soil moisture in the landscape. Both ESA and NASA have launched L-band radiometers, in the form of the SMOS and SMAP satellites respectively, to monitor soil moisture globally, every 3-day at about 40 km resolution. However, their coarse scale restricts the range of applications. While SMAP included an L-band radar to downscale the radiometer soil moisture to 9 km, the radar failed after 3 months and this initial approach is not applicable to developing a consistent long term soil moisture product across the two missions anymore. Existing optical-, radiometer-, and oversampling-based downscaling methods could be an alternative to the radar-based approach for delivering such data. Nevertheless, retrieval of a consistent high resolution soil moisture product remains a challenge, and there has been no comprehensive inter-comparison of the alternate approaches. This research undertakes an assessment of the different downscaling approaches using the SMAPEx-4 field campaign data.

## 1. INTRODUCTION

Continuous monitoring of soil moisture as a hydrological variable assists in quantifying exchanges of water, energy and carbon fluxes between the land surface and the atmosphere [1, 2]. Most importantly, remote sensing can capture spatial and temporal patterns of soil moisture in the landscape. Particularly useful are passive microwave observations at L-band because of their high sensitivity to soil moisture and favorable signal-to-noise ratio [2]. The European Space Agency (ESA) as well as the National Aeronautics and Space Administration (NASA) have launched L-band passive microwave instruments on board the Soil Moisture and Ocean Salinity (SMOS) and Soil

Moisture Active Passive (SMAP) satellites in 2009 and 2015, respectively. However, these satellites monitor global surface soil moisture at approximately 40 km spatial resolution, and this coarse spatial scale restricts the range of applications to hydro-climatological studies. Development of a consistent time series of soil moisture maps at moderate spatial resolution (1-10 km) is vital for SMOS and SMAP soil moisture to be used across a wider range of applications [3]. While various disaggregation techniques have been proposed to improve the spatial scale of the SMOS and SMAP derived soil moisture, until now there has been no rigorous test to assess the relative strengths and weaknesses of their performance by comparing against each other and a reference data set. This paper aims to evaluate the above downscaling methodologies, which could be used to produce a consistent time series of multi-sensor soil moisture retrievals at resolution equal to or finer than 10 km, for a typical Australian landscape.

## 2. STUDY AREA AND REFERENCE DATA

The Soil Moisture Active Passive Experiment-4 (SMAPEx-4) campaign was designed to contribute to the calibration and validation of SMAP soil moisture products. This experiment was carried out from 1$^{st}$ to 22$^{nd}$ of May 2015 in the semi-arid agricultural area of Yanco (Figure 1), located in the Murrumbidgee catchment in south-east of Australia. The Yanco area contains a network of soil moisture monitoring stations (as part of OzNet [4]: http://www.oznet.org.au/). During SMAPEx-4, airborne L-band passive microwave observations were also collected by the Polarimetric L-band Multibeam Radiometer (PLMR) sensor at 1-km resolution, concurrent with the overpass of SMAP and SMOS. The aircraft soil moisture retrievals, validated against ground-based measurements, have been used to spatially validate downscaled satellite retrievals. The *in situ* acquisitions from OzNet were also used for validation of downscaled soil moisture maps.

## 3. METHODOLOGY

A range of different optical-, radar-, radiometer-, and oversampling-based downscaling methods has been benchmarked with SMOS and SMAP coarse passive microwave observations to comprehensively evaluate their performance. This inter-comparison (Figure 2) was undertaken over the Yanco region using the SMAPEx-4 field campaign data, the PLMR-derived 1-km resolution soil moisture maps, as well as OzNet *in situ* soil moisture measurements as validation references.

Two types of optical-based downscaling approaches were considered for downscaling SMOS: i) physical based method, namely Disaggregation based on Physical And Theoretical scale Change (DisPATCh; [5]), and ii) semi-empirical approach explained in [6] and [7]. Downscaled

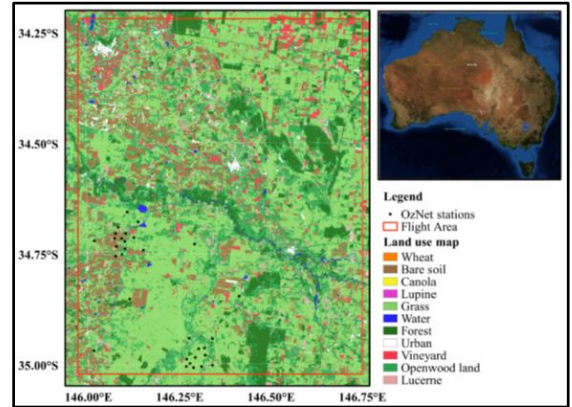

Figure 1 – The SMAPEx-4 land-use map showing aircraft sampling area of 71 km × 85 km as well as OzNet soil moisture measurement stations.

SMAP soil moisture retrievals were also produced using the radiometer-based Smoothing Filter-based Intensity (SFIM) Modulation model developed by [8], radar-based downscaling using the baseline active-passive method of SMAP [9] and the Multi-Objective Evolutionary Algorithm (MOEA) by [10], as well as an oversampling-based technique using the Backus-Gilbert method [11].

The DisPATCh model utilizes the SMOS L3 radiometric soil moisture (SMOS operational, version 2.8 obtained from the Centre Aval de Traitement des Données SMOS (CATDS) website) as background, and high resolution approximation of the Soil Evaporative Efficiency (SEE) – produced using the MODerate resolution Imaging Spectroradiometer (MODIS) soil temperature and vegetation index data – to downscale soil moisture to a resolution of 1 km. This technique was applied to the ascending and descending soil moisture observations resulting in two DisPATCh products, the DisPATChD (descending) and DisPATChA (ascending).

The semi-empirical method calibrates a downscaling model by relating the ensemble of satellite-derived brightness temperature (Tb), Normalized Difference Vegetation Index (NDVI) and Land Surface Temperature (LST) to soil moisture. This method was applied to the daily SMOS L3 products produced by the Barcelona Expert Center (BEC) and the 16-day composite of NDVI and daily LST derived from the MODIS observations to deliver ascending and descending BEC L4 data (BEC L4A and BEC L4D, respectively).

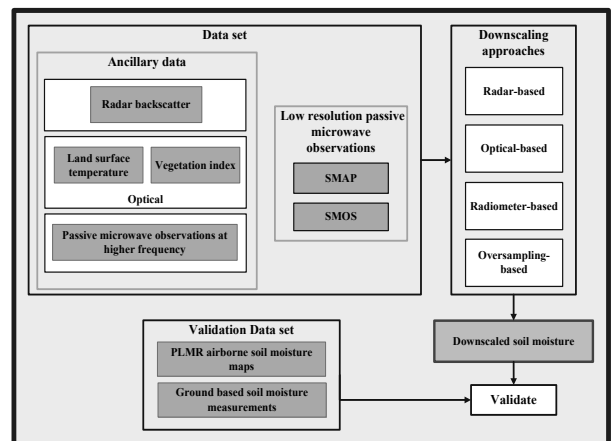

Figure 2 – Schematic of the designed procedure for multi-sensor soil moisture retrieval at medium resolution.

In the SFIM procedure, SMAP L2 Tb (SPL2SMP) is downscaled to the resolution of the Advanced Microwave Scanning Radiometer-Earth Observing System (AMSR2) Ka- band observations (~10 km), using a weighting factor. This weighting factor was derived from a ratio between the Ka-band Tb value for each grid cell at approximately 10 km and the average of Ka-band Tb across the coarse scale of SMAP Tb observations. Soil moisture retrieval is then derived from downscaled Tb through application of the Land Parameter Retrieval Model (LPRM; [12]).

The SMAP active-passive baseline algorithm downscales the SMAP L2 Tb (SPL2SMP) using a linear correlation between SMAP L2 Radar backscatter (σ°, SPL2SMA) and Tb. Soil moisture is then estimated by applying the radiative transfer model (single channel algorithm, [13]) to the downscaled Tb. These estimates are available at the NASA National Snow and Ice Data Center Distributed Active Archive Center (NSIDC DAAC) website as SMAP L2 Radar/Radiometer Half-Orbit 9 km EASE-Grid Soil Moisture, Version 4 (SPL2SMAP).

The core of the MOEA technique is a joint cost function which optimizes accuracy and spatial scale of soil moisture retrieval through giving more weight to σ°. It is important to note that MOEA technique was applied to the SMAP Tb (SPL2SMP) and σ° (SPL2SMA) pairs. The Backus-Gilbert (BG) is a theory of interpolation which provides Tb at 9 km by averaging Tb centered near a particular radius with a relatively short length of intervals. Soil moisture is then derived by applying a radiative transfer model to the downscaled Tb. This technique was applied to the SMAP morning and afternoon Tb (SPL2SMP) products and resulted in two series of products, the EnhancedD and EnhancedA, respectively.

## 4. RESULTS

Soil moisture disaggregation techniques were evaluated using soil moisture maps generated by upscaling PLMR soil moisture footprints to the scale of the disaggregated products. This upscaling helps to diminish disparity of spatial scales between the PLMR and satellite based soil moisture retrievals. The quantitative results of the validation study are currently being analyzed and will be presented, while sample of qualitative time series of downscaled and observed soil moisture maps are shown in Figure 3. This figure shows the performance of the downscaling techniques in capturing the spatio-temporal variability of soil moisture.

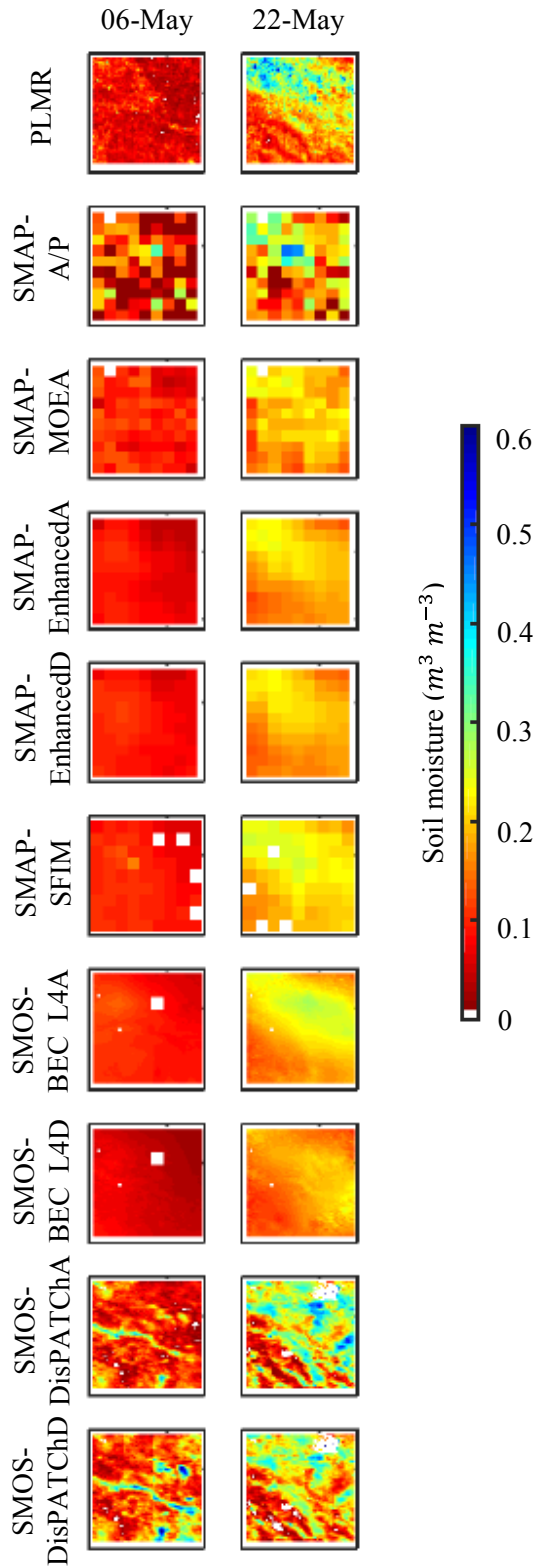

Figure 3 – Sample of the time series plots of PLMR observed reference soil moisture estimates and the range of downscaled soil moisture estimates for the Yanco region.

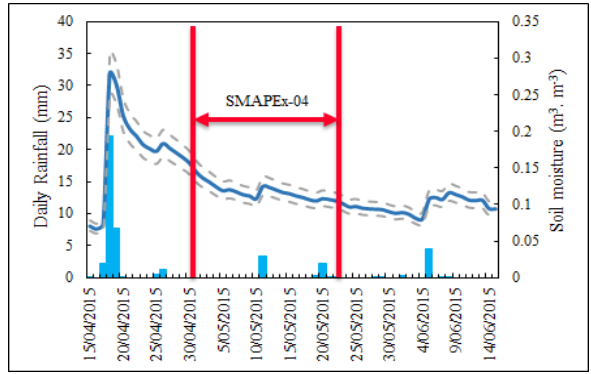

Figure 4 – Time series of *in situ* soil moisture and rainfall for the SMAPEx4 period.

The PLMR soil moisture estimates at 1 km have consistency with the occurrence of precipitation events and mimic dry down cycles (see Figure 3 and Figure 4). There is no clear evidence to show any downscaling technique is superior for disaggregation of SMOS and/or SMAP, but it is obvious that among the downscaled products, DisPATCh revealed better agreement with the spatial and temporal pattern of PLMR soil moisture than other products. However, it could not deliver any soil moisture under cloudy skies because optical imagery, which is the key component in the DisPATCh downscaling process, is not available under cloud coverage. This is unlike microwave-based approaches that are capable of downscaling soil moisture under all-weather conditions. This capability is due to transparency of microwave observations to the opaque cloud cover.

## 5. CONCLUSION AND RECOMMENDATION

This paper presents a qualitative analysis of the different soil moisture downscaling technique performances, to overview their applicability for disaggregation of SMOS and/or SMAP. This preliminary assessment is the first stage of a comprehensive analysis of alternative downscaled soil moisture products. While cloudy skies limit the applicability of optical-based downscaling techniques at global scale, the use of geostationary based optical sensors, which collect optical data at about 30 minute time intervals, may help to overcome this issue by increasing the chance of capturing cloud-free observations. Alternatively, microwave-based approaches are required to avoid this problem.

## 6. ACKNOWLEDGEMENT

This study has been conducted within the framework of an ARC Discovery Project (MoistureMonitor, DP140100572) and Infrastructure grant (LE0453434). Monash University is acknowledged for awarding a postgraduate scholarship to Sabah Sabaghy to pursue her PhD.